\documentclass[12pt]{article}
\usepackage{slashed}
\usepackage{amsmath}
\usepackage{amssymb}
\usepackage{amsthm}
\usepackage{psfrag}
\usepackage{graphicx}

\newcommand\bw{\begin{widetext}}
\newcommand\ew{\end{widetext}}
\newcommand{\be}{\begin{equation}}
\newcommand{\ee}{\end{equation}}
\newcommand{\beqa}{\begin{eqnarray}}
\newcommand{\eeqa}{\end{eqnarray}}

\newcommand\m{\mu}
\newcommand\D{\Delta}
\newcommand\n{\nu}

\renewcommand\a{\alpha}
\renewcommand\b{\beta}

\def\e{{\rm e}}
\def\d{\partial}
\newcommand{\bseq}{\begin{subequations}}
\newcommand{\eseq}{\end{subequations}}

\renewcommand{\L}{\Lambda}

\begin{document}

\title{\vspace{-2cm} 
\begin{flushright}
{\normalsize
CERN-PH-TH-2015-112} \\
\vspace{-0.5cm}
{\normalsize INR-TH/2015-011}
\end{flushright}
\vspace{0.5cm} 
{\bf Emergent Lorentz invariance with\\ chiral fermions}}

\author{Ivan Kharuk$^{1,2}$,  Sergey Sibiryakov$^{1,3,4}$\\[2mm]
{\normalsize\it $^1$ Institute for Nuclear Research of the
Russian Academy of Sciences,}\\[-1mm]
{\normalsize\it 60th October Anniversary Prospect, 7a, 117312
Moscow, Russia}\\
{\normalsize\it $^2$ Moscow Institute of Physics and
  Technology,}\\[-1mm]
{\normalsize\it Institutskii per. 9, Dolgoprudny 141700, Moscow
  Region, Russia}\\
{\normalsize\it  $^3$ CERN Theory Division, 
CH-1211 Geneva 23, Switzerland}\\
{\normalsize\it $^4$Institut de Th\'eorie des Ph\'enom\`enes Physiques, 
 EPFL, CH-1015 Lausanne, Switzerland}}
\date{}
\maketitle

\begin{abstract}
We study renormalization group flows in strongly interacting field
theories with fermions that correspond to transitions between a theory
without Lorentz invariance at high energies down to a theory with approximate
Lorentz symmetry in the infrared. Holographic description of the
strong coupling is used. The emphasis is made on emergence of
chiral fermions in the low-energy theory.
\end{abstract}

\section{Introduction}
\label{sec:1}
Recent years have witnessed a revival of the idea
\cite{Nielsen:1978is,Chadha:1982qq} that Lorentz invariance (LI),
instead of being a fundamental property of Nature, may emerge only as an
approximate symmetry of the low-energy physics. An important
motivation to explore this possibility comes from the proposal by P.~Ho\v rava 
\cite{Horava:2009uw} that abandoning LI improves the ultraviolet
behavior of gravity. In this proposal, departure from LI at large
energies allows to postulate anisotropic (Lifshitz) scaling \cite{Lifshitz} which
renders the gravity theory power-counting renormalizable. There
is a version of Ho\v rava--Lifshitz gravity \cite{Blas:2009qj,Blas:2010hb}
which is viable and
passes the existing experimental constraints \cite{Blas:2014aca}. On
the phenomenological side, breaking of LI has been invoked in the
construction of interesting long-distance modifications of gravity 
\cite{Jacobson:2000xp,ArkaniHamed:2003uy,Dubovsky:2004sg}. These
provide a playground for the analysis of possible deviations from
general relativity and may be relevant for cosmology
\cite{Rubakov:2008nh,Clifton:2011jh}.

Emergence of an effective Lorentz symmetry at low energies is a fairly
common phenomenon in condensed matter physics, see
e.g. \cite{CM1,CM2,Lee:2006if,Herbut:2009qb,Giuliani:2011dc,CM3}. Also,
it is not hard to
construct particle physics models with this property. Indeed, the
original works \cite{Nielsen:1978is,Chadha:1982qq} considered
Yang--Mills theory and quantum electrodynamics with an addition of
Lorentz violating terms of dimension four\footnote{Lorentz violating
  terms of dimension two cannot be constructed from the fields of the theory and those of dimension 
three can be forbidden by imposing the discrete $CPT$ symmetry.} 
in the Lagrangian
and showed that the coefficients of those terms vanish along the
renormalization group (RG) flow towards the infrared (IR). This result
persists in the Lorentz violating extension of the full Standard Model
\cite{Giudice:2010zb}. Furthermore, a general argument
\cite{Sundrum:2011ic,Bednik:2013nxa} shows that LI fixed points of RG
flows are IR stable with respect to $CPT$-invariant deformations 
and thus have non-empty basin of attraction in the
space of $CPT$-invariant 
non-relativistic theories. All theories belonging to this
basin of attraction exhibit emergent LI\footnote{Of course, this does
  not imply that {\em any} non-relativistic theory becomes LI in the
  infrared; for example, there are RG flows terminating at fixed points with Lifshitz scaling.}. In two space-time
dimensions it has been shown \cite{Sibiryakov:2014qba} 
that under broad assumptions interacting
fixed points with {\em isotropic} scaling are necessarily LI. 

However, this  
scenario faces an important
challenge when confronted with the precision tests of LI in particle
physics experiments
\cite{Mattingly:2005re,Kostelecky:2008ts,Liberati:2013xla}. 
In a weakly coupled theory the running of the coefficients in front of
marginal Lorentz violating operators is logarithmic. If they are of
order one\footnote{In principle, one could consider a setup where at
  tree level LI is an exact symmetry of the matter sector and is
  violated only in gravity \cite{Pospelov:2010mp}. Then, Lorentz
  violation is transmitted to matter only through loops with gravitons
which are suppressed. However, the departure from the
homogeneous Lifshitz scaling in the UV required by
 this approach can compromise the
high-energy properties of the theory \cite{Colombo:2014lta}.} 
at a certain ultraviolet (UV) scale $\L_*$, say, of order Planck mass,
they become only mildly suppressed at the experimentally accessible
energies. These operators would, in particular, modify the dispersion
relations of elementary particles producing species-dependent shifts
of their maximal velocities \cite{Coleman:1998ti}. Experimental
constraints on this effect are very tight \cite{Kostelecky:2008ts} and
the logarithmic suppression is by far insufficient to satisfy
them. Higher-dimension Lorentz violating operators are suppressed by $\L_*$ and are less problematic. Still, important bounds on their contributions into particles' dispersion relations
can be obtained from astrophysics
\cite{Jacobson:2002hd,Galaverni:2007tq,Maccione:2008iw,
Liberati:2012jf,Rubtsov:2013wwa}. 

There are two ways to address this challenge. One is to stay within
perturbative regime and assume that the theory has another symmetry 
which entails LI as an accidental symmetry at
low energies. Remarkably, the role of the extra symmetry can be played
by non-relativistic supersymmetry
\cite{GrootNibbelink:2004za,Pujolas:2011sk}. Of course, in a realistic
model the supersymmetry must be broken which leads to the generation
of potentially dangerous Lorentz violating operators. However, if the scale of
supersymmetry breaking is hierarchically lower than $\L_*$, the
coefficients of these operators are strongly suppressed and can be
compatible with the experimental bounds. It should be mentioned though
that it is unclear if the required version of supersymmetry is
compatible with Lifshitz scaling at high energy
\cite{Redigolo:2011bv}.   

\begin{figure}[t]
\begin{center}
\includegraphics[width=0.55\textwidth]{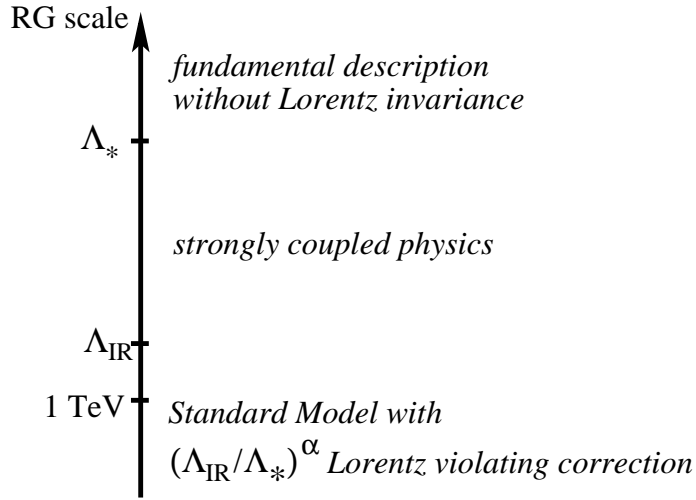}
\caption{Various scales in the proposal for 
emergence of Lorentz invariance due to strong
 coupling. Above $\Lambda_*$ the theory is essentially
 non-relativistic. The strongly coupled RG flow below $\Lambda_*$
 drives the theory towards a Lorentz invariant fixed point. 
A relevant deformation terminates the flow at $\Lambda_{IR}$ where
the theory enters into a
 confining phase and the Standard Model fields emerge
 as composite states.
\label{fig:1}}
\end{center}
\end{figure}

Another approach was proposed in 
\cite{Bednik:2013nxa} (see also \cite{Kiritsis:2012ta}) and is
illustrated in Fig.~\ref{fig:1}. It is  
based on the observation that in
strongly interacting theories the RG flow (understood in the Wilsonian
sense) is fast and generally results in a power-law dependence of the
couplings on the RG scale. Thus, if the Standard Model is embedded
into a theory that passes through a strongly coupled RG evolution
between the Lorentz violation scale $\L_*$ and some lower scale
$\L_{IR}$ (still above the experimentally accessible energies), the
coefficients of the Lorentz violating operators are suppressed by
$(\L_{IR}/\L_*)^\a$ with $\a>0$. The latter exponent is
controlled by the dimension of the Least Irrelevant Lorentz Violating
Operator (LILVO) \cite{Bednik:2013nxa}. If $\a$ and the hierarchy
between the scales $\L_{IR}$, $\L_*$ are large enough, the
experimental constraints will be satisfied. 
Note that in this scenario it is sufficient that strong coupling 
occurs only in the matter sector and therefore is described, at least
in principle, by the standard methods of quantum field theory. 
Gravity can remain weakly coupled at all energies.   

It is worth stressing that this scenario implies that the whole SM
sector passes through the regime of strong coupling. In other words,
all SM fields must emerge as bound states of the strong dynamics with
the compositeness scale set by $\Lambda_{IR}$. Construction of
realistic models with this property is challenging. A useful tool is
provided by the gauge/gravity (holographic) 
correspondence that enables to describe 4-dimensional strong dynamics
in terms of weakly coupled theory in
one dimension higher. This approach was used in~\cite{Bednik:2013nxa} to 
implement the above mechanism for emergence of LI
in a scalar toy model. The purpose of this work is
to extend the analysis to theories with fermions. We focus on
the case when the non-relativistic UV theory is
invariant under spatial rotations, so fermions are defined as fields
transforming in the spinor representation\footnote{We shall assume
  that these fields obey anti-commutation relations.} 
of $SO(3)$ (or
$SO(d-1)$ in the general case of $(d-1)$ spatial dimensions).  

Inclusion of fermions in the framework of emergent LI brings 
the following puzzle. The key role   
 in the structure of the Standard Model is played by {\em chiral}
 fermions. 
However, the very definition of chirality relies on the existence of
the Weyl representations of the Lorentz group. How chiral fermions can
appear in a theory that fundamentally does not possess LI~? We will
see that the answer to this question relies on a simple
kinematic property:
an $SO(3)$
spinor of the UV theory has two independent 
components which matches precisely the number of components of 
a four-dimensional Weyl spinor\footnote{The match holds for
  any
even-dimensional space-time.}. The degrees of freedom of the UV
fermion are dressed by the strongly coupled RG evolution and at low
energy constitute a chiral fermion. A necessary and sufficient
requirement for this to happen is presence of a gapless mode in the
low-energy theory. We will present holographic models of RG
flows where this is indeed the case. These constructions 
will enable to compute explicitly 
the LV
contributions in the low-energy theory 
and verify their 
power-law suppression.

The paper is organized as follows. In Sec.~\ref{sec:2} we briefly review the
holographic formalism for fermions in the
standard relativistic setting. In Sec.~\ref{sec:3} we consider a
simplified model
describing 
a non-relativistic fermion coupled to a strongly interacting
relativistic sector. The latter is taken to be vector-like, i.e. 
it does not contain any massless chiral bound states. We
show that this construction leads to appearance of a gapless
chiral mode in the low-energy theory together with emergent LI. 
In Sec.~\ref{sec:4} we turn to a setup modeling an RG flow
from a theory with Lifshitz scaling 
in UV towards a LI infrared fixed point. We assume that the 
RG flow is stopped by the theory entering into a confining
phase. We show that a suitable pattern of confinement again leads to
emergence of a chiral bound state. 
In Sec.~\ref{app:1} we discuss emergence of LI in theories with
fermions 
in odd
space-time dimensions.
Section \ref{sec:5} is devoted to conclusions.

\section{Holography for relativistic fermions}
\label{sec:2}
In this section we review the holographic correspondence for fermions in
the LI case following
\cite{Henningson:1998cd,Mueck:1998iz,Henneaux:1998ch,Iqbal:2009fd}. 
It establishes a relation between a strongly coupled conformal field
theory (CFT) with large number of degrees of freedom in $d$ dimensions
and gravity in $(d+1)$-dimensional anti-de Sitter (AdS)
space-time. The latter has the metric,  
\begin{equation} 
\label{eq:metr}
ds^2 = \left( \frac{l}{u} \right) ^2 \big(-dt^2+dx^idx^i+du^2\big)~,
~~~~i=1,\ldots, d-1,
\end{equation}
where $l$ is the AdS radius. A fermionic operator ${\cal
  O}_\psi$ in the CFT corresponds to a fermion $\psi$ in AdS. The
action for a free fermion with mass $M$ reads,
\begin{equation} 
\label{BulkAction}
S= - \int d^{d+1}x \sqrt{|g|}\; i \big(\bar{\psi}{\slashed{D}}\psi
 - M\bar{\psi}\psi\big)+S_{\d}\;,
\end{equation}
where $\bar\psi=\psi^\dagger \Gamma^t$ and the Dirac operator
is\footnote{We use capital Latin letters from the beginning of the
alphabet $A,B,\ldots$ for indices in the local Lorentz frame in $d+1$
dimensions; capital letters from the middle of the alphabet
$M,N,\ldots$ are used for $(d+1)$-dimensional space-time indices;
Greek letters $\m,\n,\ldots$ denote indices in $d$ dimensions spanned
by $t$ and $x^i$, $i=1,\dots,d-1$ 
(as this space-time is flat, we do not distinguish
the tangent-space indices and the indices in the local Lorentz frame).}
\be
\label{dirac}
\slashed D=e^M_A \Gamma^A D_M~,~~~~
D_M=\bigg(\d_M+\frac{1}{2}\omega_{ABM}\Sigma^{AB}\bigg)~,~~~~~
\Sigma^{AB}=\frac{1}{4}[\Gamma^A,\Gamma^B]\;;
\ee
$\Gamma^A$ are the $(d+1)$-dimensional Dirac matrices.
The boundary term $S_{\d}$ is needed to ensure the correct variational
principle and will be specified shortly. The sign of the action is fixed
by unitarity \cite{Iqbal:2009fd}. Choosing the
diagonal vielbein and computing the corresponding spin connection,
\be
\label{vielbein}
e^M_A=(u/l)\, \delta^M_A~,~~~~\omega_{u\b\m}=\eta_{\b\m}/u\;,
\ee
with all other components of $\omega$ vanishing, one 
simplifies the expression for the Dirac operator,
\be
\slashed D=\frac{u}{l} \big(\Gamma^u\d_u+\Gamma^\m\d_\m\big)-\frac{d}{2l}\Gamma^u\;.
\ee
To solve the Dirac equation
\be
\label{bulkDir}
(\slashed D-M)\psi=0
\ee 
one decomposes the spinor into eigenvectors with respect to $\Gamma^u$,
\be
\label{decomp}
\psi=\psi_++\psi_-~,~~~~~ \Gamma^u\psi_\pm=\pm \psi_\pm\;.
\ee
Performing the Fourier transform along the $d$ flat dimensions,
$\psi\propto \e^{ip_\m x^\m}$, we obtain the general solution,
\bseq
\label{pmsol}
\begin{align}
&\psi_-(\vec{p},u)=(pu)^{(d+1)/2}\big(\chi_{1}(\vec{p})J_{Ml+1/2}(pu)
+\chi_{2}(\vec{p})Y_{Ml+1/2}(pu)\big)\;,
\label{pmsol2}\\
&\psi_+(\vec{p},u)=(pu)^{(d+1)/2}\frac{ip_\m\Gamma^\m}{p}
\big(\chi_{1}(\vec{p})J_{Ml-1/2}(pu)
+\chi_{2}(\vec{p})Y_{Ml-1/2}(pu)\big)\;,
\label{pmsol1}
\end{align}
\eseq
where we have introduced $p\equiv \sqrt{-p_\m p^\m}$. 

For $Ml>1/2$ the
mode associated with $\chi_{2}$ is not normalizable.
It is
interpreted as a source for the operator ${\cal O}_\psi$ in the CFT. More precisely, the source $\chi$ is defined by the relation
\be
\label{sourcedef}
\chi=\lim_{u\to 0}(u/l)^{Ml-d/2}\psi_-\;.
\ee 
The partition function,
\be
\label{partition}
Z[\chi]\equiv\int d\psi d\bar\psi~ \e^{iS_{AdS}}\;,
\ee
with the integral taken over configurations satisfying the
boundary condition\footnote{Generally, the integration in
  (\ref{partition}) should be performed over all fields present in the
AdS theory, including the metric, subject to appropriate boundary
conditions. For simplicity, we omit them in our discussion.} 
(\ref{sourcedef}) provides the generating
functional for the correlators of ${\cal O}_\psi$.
The
CFT without sources corresponds to $\chi_2=0$. 
The dimension of ${\cal O}_\psi$ is then read off from the small-$u$
behavior of the component 
$\psi_+$:
\be
\label{dimO}
\psi_+\propto \frac{ip_\m\Gamma^\m}{p}\chi_1(p) (pu)^{Ml+d/2}
~~~~\Longrightarrow~~~~
\dim {\cal O}_\psi=Ml+\frac{d}{2}\;.
\ee
At $Ml<-1/2$ the situation is reversed. Now the
non-normalizable mode is associated with $\chi_1$ which dominantly
contributes into the component $\psi_+$. Thus, the latter is
interpreted as a source, whereas $\psi_-$ determines the dimension of
the dual operator,
\be
\label{dimOtilde}
\dim {\cal O}'_\psi=-Ml+\frac{d}{2}\;.
\ee
In the window 
$-1/2<Ml<1/2$ both $\chi_1$ and $\chi_2$ modes are normalizable and
the system admits two different quantizations
\cite{Iqbal:2009fd}. In this paper we restrict to the range
$Ml>-1/2$ and choose the quantization with the source related to
$\psi_-$ by Eq.~(\ref{sourcedef}); the dimension of the operator is given
by (\ref{dimO}). Note that at $Ml=-1/2$ the expression (\ref{dimO})
saturates the lower unitarity bound on the dimension of a spin-half operator in
$d$-dimensional space-time \cite{dim1,Mack:1975je,Minwalla:1997ka}.

The variation of the total
action (\ref{BulkAction}) must vanish on the solution. Integrating the
bulk term by parts and setting the contribution at $u=\infty$ to zero 
we obtain,
\be
\label{Svar}
\delta S=
\lim_{\epsilon\to 0}i\int_{u=\epsilon} d^d
x\;\big(\epsilon/l\big)^{-\frac{d-1}{2}}
(\bar\psi_- \delta\psi_+-\bar\psi_+\delta\psi_-)
+\delta S_\d\;.
\ee
The component $\psi_-$ is subject to the Dirichlet boundary
condition (\ref{sourcedef}), 
so its variation vanishes. However, $\delta\psi_+$ cannot
be set to zero. To cancel this contribution one chooses 
\cite{Henneaux:1998ch}
\be
\label{sboundary}
S_\d=-\lim_{\epsilon\to 0}i\int_{u=\epsilon} d^d
x\;\big(\epsilon/l\big)^{-\frac{d-1}{2}}\;
\bar\psi_-\psi_+\;.
\ee
If the AdS space is bounded by a brane at finite $\epsilon$, which
corresponds to a finite UV cutoff in the dual CFT, the term
(\ref{sboundary}) must be evaluated on this UV brane.

We assume that the CFT enters into a confining phase at low energy, so
that it possesses a discrete spectrum of excitations, cf.~\cite{HF}. 
This is realized by placing a brane\footnote{To solve the Einstein
  equations, the brane must have negative tension. This does not lead
  to pathologies if the gravitational field obeys suitable
  boundary conditions, see e.g.~\cite{RS}.} 
at $u=L$ (referred to as ``IR
brane'' below) 
and removing the region
$u>L$. 
Physically, one interprets $L^{-1}$ as the confining scale.
We will consider two choices of boundary conditions on the
IR brane, which we call
${\cal B}_+$ and ${\cal B}_-$. In the ${\cal B}_+$ case one imposes
vanishing of $\psi_+$ on the brane,
\be
\label{Bplus}
{\cal B}_+:~~\psi_+\big|_{u=L}=0\;, 
\ee
while the $\psi_-$ component is arbitrary.
In the case ${\cal B}_-$ the situation is opposite,
\be
\label{Bminus}
{\cal B}_-:~~\psi_-\big|_{u=L}=0\;,
~~~~~~~\psi_+\big|_{u=L}~ \text{--- arbitrary}\;.
\ee 
The spectrum of masses is
obtained by substituting (\ref{pmsol}) into Eq.~(\ref{Bplus}) or
(\ref{Bminus})  (recall that $\chi_{2}=0$ due to the
boundary condition at $u\to 0$),
\be
\label{masses}
m^{(\pm)}_j=L^{-1}\m_j^{(Ml\mp 1/2)}\;,
\ee
where $\m_j^{(\nu)}$ is the $j$-th positive root of the Bessel
function. Massless modes should be studied separately. A
straightforward analysis shows that massless modes are absent in
the case ${\cal B}_+$. On the other hand, in the case
${\cal B}_-$ there is a massless mode of the form,
\be
\label{massless}
\psi_+=\chi_{0} (u/l)^{Ml+d/2}~,~~~~\psi_-=0\;,
\ee 
where the spinor $\chi_{0}$ satisfies the equation $p_\m\Gamma^\m\chi_{0}=0$.

Henceforth we specify
to the case of even $d$; we will return to the odd-$d$ case in
Sec.~\ref{app:1}. It is convenient to choose
the $(d+1)$-dimensional
gamma matrices as
\be
\label{gammaeven}
\Gamma^\m=\gamma^\m~,~~~\m=0,\ldots,d-1~,~~~~~~~\Gamma^u=\gamma^{d+1}\;,
\ee
where $\gamma^\m$ are $d$-dimensional Dirac matrices and $\gamma^{d+1}$ is
the chirality matrix ---
the $d$-di\-men\-sio\-nal analog of $\gamma^5$. 
From (\ref{sourcedef}) we see that the source $\chi$ and the
corresponding operator ${\cal O}_\psi$ belong to the Weyl
representation of the $d$-dimensional Lorentz group. However, the
spectrum of the bound states contains a chiral massless mode only in
the case ${\cal B}_-$. In the case ${\cal B}_+$ all eigenmodes are
massive and have both left and right components combining into the
full Dirac representation. In this sense the dual low-energy theory arising in
the ${\cal B}_+$ case is vector-like.

\section{Chiral modes from non-relativistic fermions}
\label{sec:3}

We want to couple the (deformed) CFT of the previous section to a
non-relativistic fermion and study the resulting RG flow. To this end
we introduce a UV cutoff $\Lambda_*$ and at this scale promote the
source $\chi$ to an elementary dynamical field with Lorentz violating
action $S_\chi$. 
The total action at the scale $\Lambda_*$ takes the form,
\be
\label{Stot}
S=S_{CFT}+S_{\chi}+\Lambda_*^{-Ml}\int d^dx\, \bar\chi\, {\cal O}_\psi\;.
\ee
We will assume $S_\chi$ to be invariant under spatial
rotations\footnote{It is worth noting, however, that the analysis of this
section can be straightforwardly generalized to the case without
spatial isotropy.}. 
For concreteness, let us start from a simple choice,
\be
\label{Schi}
S_\chi=-b\int d^dx\,
i(\bar\chi\gamma^0\d_0\chi+v\,\bar\chi\gamma^i\d_i\chi)\;,
\ee
where the velocity $v$ of the $\chi$-field is different from unity.
The parameter $b$ with dimension of length has been introduced to
render the action dimensionless.
Note that we have not included the
Lorentz violating operator without derivatives 
$\bar\chi\gamma^0\chi$ which is odd under the
charge-conjugation (and $CPT$). For the moment we will assume these
symmetries, though they are not essential for the present setup (see
below).  

On the AdS side the above setup is realized by cutting the space-time
with a UV brane at $u=\Lambda_*^{-1}$ and supplementing the action
(\ref{BulkAction}) with the boundary term,
\be
\label{SUV}
S_{UV}=-b\int_{u=\Lambda_*^{-1}} d^dx\,
i(\bar\psi_-\gamma^0\d_0\psi_-+v\,\bar\psi_-\gamma^i\d_i\psi_-)\;.
\ee
In this way we obtain a theory in the slice
of AdS bounded by a UV and IR branes, cf.~\cite{RS,HF}.
Without loss of generality one can identify $\Lambda_*^{-1}$ with the
AdS radius $l$; this is done from now~on.

The canonical dimension of the elementary fermion $\chi$ is
$(d-1)/2$. Recalling the dimension (\ref{dimO}) of the operator 
${\cal O}_\psi$ we observe that the interaction between $\chi$ and 
${\cal O}_\psi$ in (\ref{Stot}) has dimension 
$d+Ml-1/2$. If $Ml>1/2$ this interaction is irrelevant and the
Lorentz violating sector simply decouples from the CFT in the
infrared. Hence, no LI emerges in this case, similar to what happens
in the scalar model at irrelevant coupling \cite{Bednik:2013nxa}.
In what follows
we restrict to the range
\be
\label{Minterest}
-1/2<Ml<1/2\;,
\ee
when the interaction is
relevant and generates a strongly coupled RG flow.
Furthermore, we will focus on the choice ${\cal B}_+$
of the 
IR boundary
conditions. 
As discussed in the previous section, the CFT in
this case does not possess any massless modes. Then the total system
has a single gapless mode coming from the elementary fermion
$\chi$ which, as we are going to see shortly,  
survives down to IR. On the other hand, in the case of boundary
conditions
${\cal B}_-$ the total number of gapless modes is 2 (one from the CFT
and one from $\chi$).
It is straightforward to show that they 
pair and form
a gap \cite{HF}. As our aim is to study the emergence of chirality,
this case is not of interest to us.

From (\ref{BulkAction}), (\ref{SUV}) one derives the boundary
condition on the UV brane,
\be
\label{BCUV}
\big[b(\gamma^0\d_0+v\,\gamma^i\d_i)\psi_-+\psi_+\big]\big|_{u=l}=0\;.
\ee
The spectrum of eigenmodes is determined by imposing (\ref{BCUV})
together with the IR boundary condition (\ref{Bplus}) on the general
bulk solution (\ref{pmsol}). This leads to the wavefunctions,
\be
\label{psibulk}
\psi_-=\chi_{1}f_-(u)~,~~~~\psi_+=\frac{ip_\m\gamma^\m}{p}\chi_{1}f_+(u)\;,
\ee
with 
\be
\label{functions}
f_{\pm}(u)=(pu)^{(d+1)/2}\bigg[J_{Ml\mp 1/2}(pu)-
\frac{J_{Ml-1/2}(pL)}{Y_{Ml-1/2}(pL)}Y_{Ml\mp 1/2}(pu)\bigg]\;
\ee
and the spinor $\chi_1$ obeying the relation,
\be
\label{BCUV1}
\bigg[p_0\gamma^0\bigg(1+p\,b\,\frac{f_-(l)}{f_+(l)}\bigg)
+p_i\gamma^i\bigg(1+v\,p\,b\,\frac{f_-(l)}{f_+(l)}\bigg)\bigg]\chi_{1}=0\;.
\ee
We are interested in the gapless mode whose $d$-dimensional 
momentum squared is 
much smaller than the confinement scale, 
\be
\label{light}
pl\ll pL\ll 1\;.
\ee
Upon expanding the ratio $f_-(l)/f_+(l)$ under these assumptions
Eq.~(\ref{BCUV1}) simplifies, 
\be
\label{BCUV2}
\bigg[-\omega\gamma^0\bigg(1
+(1-2Ml)\frac{b}{l}\bigg(\frac{l}{L}\bigg)^{1-2Ml}\bigg)
+k_i\gamma^i\bigg(1
+v\,(1-2Ml)\frac{b}{l}\bigg(\frac{l}{L}\bigg)^{1-2Ml}\bigg)\bigg]\chi_{1}=0\;,
\ee
where we denoted $\omega\equiv -p_0$, $k_i\equiv p_i$.
It is natural to take the parameters $b$ and $l$ to be of the same
order as they both characterize the UV properties of the system. 
Then (\ref{BCUV2}) has the same form as 
the standard relativistic
equation for a Weyl spinor, up to corrections suppressed by a power of
the small
ratio $(l/L)$ of the IR and UV cutoffs. 
Note that the maximal suppression achievable in
this setup is $(l/L)^2$ which is
reached at $Ml\approx -1/2$.
Squaring the combination of the $\gamma$-matrices in (\ref{BCUV2})
yields the dispersion relation,
\be
\label{disprel1}
\omega=\pm |{\bf k}|\bigg[1
+(v-1)(1-2Ml)\frac{b}{l}\bigg(\frac{l}{L}\bigg)^{1-2Ml}\bigg]\;.
\ee
We see that the velocity of the mode is close to one for large
hierarchy between the UV and IR scales.
The equation for the amplitude of the positive-frequency mode then
takes the form,
\be
\label{BCUV3}
\Sigma_{\bf k}\chi_{1}=0~,~~~~~~
\Sigma_{\bf k}\equiv-|{\bf k}|\gamma^0+k_i\gamma^i\;.
\ee
Recalling that $\chi_1$ also satisfies the chirality condition 
$\Gamma^u\chi=\gamma^{d+1}\chi_1=-\chi_1$ 
(see (\ref{decomp}), (\ref{gammaeven})), we conclude that it 
has $2^{d/2-2}$ independent components with the same
structure as a
Weyl spinor in $d$ dimensions. In particular, in the case
$d=4$ it has a single component corresponding to the spin pointing along the
direction of the spatial
momentum. 

The above derivation was carried under the assumptions
(\ref{light}). Using (\ref{disprel1}) one finds that the rightmost
inequality is satisfied as long as the momentum of the mode does not
exceed certain upper limit, 
\be
\label{kbound}
|{\bf k}|\ll\frac{1}{L\sqrt{|v-1|}}\bigg(\frac{L}{l}\bigg)^{1/2-Ml}\;.
\ee
This bound is parametrically larger than $1/L$, but smaller
than $1/l$. To understand what happens at higher momenta we expand
$f_{\pm}(l)$ for $pl\ll 1$ treating the product $pL$ as a quantity of
order one. This yields,
\[
p\,b\frac{f_+(l)}{f_-(l)}=\frac{b}{l}\, {\cal J}(pL)\,(pl)^{1-2Ml}\;,
\]
where ${\cal J}(z)$ is a combination of Bessel functions with order-one
coefficients. The eigenvalue equation following from (\ref{BCUV1}) can
be cast into the form\footnote{In deriving (\ref{disprel22}) we have
  neglected terms on the l.h.s. suppressed by positive powers of $pl$.
This is justified only for the gapless mode which we are interested
in. For higher excitations the omitted terms are important.},
\be
\label{disprel2}
\frac{(pL)^{1+2Ml}}{{\cal J}(pL)}=
2(v-1)\frac{b}{l}\bigg(\frac{l}{L}\bigg)^{1-2Ml} {\bf k}^2 L^2\;.
\ee
Note that the l.h.s. depends only on the LI combination
$p=\sqrt{\omega^2-{\bf k}^2}$.
Inverting (\ref{disprel2}) we
obtain the dispersion relation,
\be
\label{disprel33}
\omega^2={\bf k}^2+\frac{1}{L^2}\, F\bigg[2(v-1)\frac{b}{l}
\bigg(\frac{l}{L}\bigg)^{1-2Ml} {\bf k}^2 L^2\bigg]\;,
\ee
where $F(z)$ is a dimensionless function. The dispersion
relation at low momenta is obtained by expanding $F(z)$ in Taylor
series. For the gapless mode the
zeroth-order term vanishes and the linear term gives
(\ref{disprel1}). It is instructive to consider the next term in the
expansion which gives a contribution into the dispersion relation 
of order,
\be
(v-1)^2(b/l)\, (l/L)^{-4Ml}\, l^2 {\bf k}^4\;.
\ee
For $Ml<0$ the effective scale entering this
contribution is parametrically 
higher than the UV scale $l^{-1}$. Thus, we conclude
that the strongly coupled RG flow can suppress also the higher-order
Lorentz violating corrections to the dispersion relation. For high
$|{\bf k}|$ the Taylor expansion does not apply. The analysis then
depends on the sign of $(v-1)$. If the velocity of the elementary
fermion is superluminal, $v>1$, Eq.~(\ref{disprel2}) implies
that at $|{\bf k}|\to\infty$ the combination $pL$
approaches the first positive root of the function ${\cal J}$,
\[
pL\approx z_1~,~~~~{\cal J}(z_1)=0\;.
\]
In this regime the dispersion relation again becomes
relativistic, but with a non-vanishing mass,
cf. \cite{Bednik:2013nxa}. It is unclear whether this
recovery of LI dispersion relation at high momenta 
is a peculiarity of holographic models, or is a more general
property of emergent LI in strongly coupled systems with superluminal
propagation.  
On the other hand, in the subluminal case
$v<1$ the deviation of the dispersion relation from the relativistic
form grows with $|{\bf k}|$. Both in the sub- and superluminal
cases it is easy to check that  
the spinor wavefunction satisfies (\ref{BCUV3}) for all ${\bf k}$ and
hence describes a state with a fixed helicity.

Let us return to the case of low momenta satisfying (\ref{kbound}) and
write down the effective action for the gapless
mode. This is obtained by substituting the wavefunctions
(\ref{psibulk}) into the action composed of (\ref{BulkAction}),
(\ref{sboundary}) and 
(\ref{SUV}). One finds that the bulk contribution vanishes and the
action is given by the boundary terms on the UV brane\footnote{At this
stage we do not impose the UV boundary condition (\ref{BCUV}).}, 
\be
\label{effact}
\begin{split}
S_{eff}&=\int d^dp\;
\bigg[\frac{f^*_-(l)f_+(l)}{p}\,
\bar\chi_{1}p_\m\gamma^\m\chi_{1}
+b|f_{-}(l)|^2\,\bar\chi_{1}(-\omega\gamma^0+v\,k_i\gamma^i)\chi_{1}\bigg]\\
&=-i\int d^dx\;\bigg[\bar{\hat\chi}_1\gamma^\m\d_\m\hat\chi_1
+(1-2Ml)\frac{b}{l}\bigg(\frac{l}{L}\bigg)^{1-2Ml}
\bar{\hat\chi}_1(\gamma^0\d_0+v\,\gamma^i\d_i)\hat\chi_1\bigg]\;,
\end{split}
\ee
where in the last equality we changed to the canonically normalized
field
\be
\label{chican}
\hat\chi_1=\chi_1\sqrt{\frac{f^*_-(l)f_+(l)}{p}}
\ee
and switched back to the coordinate representation. We observe that
the action consists of the LI piece describing a Weyl fermion and a
suppressed Lorentz violating contribution\footnote{For a single field
  this contribution can be removed by a rescaling of the space or time
  coordinate. However, this will no longer be possible if the system
  contains several particle species.}. Variation of this action
reproduces the equation of motion (\ref{BCUV}).

From the viewpoint of the dual field theory these results can be
understood as follows.
The coupling to the
elementary fermion generates the RG flow towards the CFT corresponding
to the alternative quantization of the bulk theory mentioned in
Sec.~\ref{sec:2}. The latter contains a fermionic operator ${\cal
  O}'_\psi$ with the dimension (\ref{dimOtilde}). The deviation from LI
is governed at low energies by the irrelevant deformation of the CFT
action, 
\be
\delta S_{CFT}\propto \int d^dx\;l^{1-2Ml} {\bar{\cal O}}'_\psi
\gamma^0\d_0{\cal O}'_\psi\;.
\ee
In the presence of confinement this deformation produces Lorentz
violating corrections to the effective action of the bound states
which depend on the ratio $l/L$ precisely in the same way as in
(\ref{effact}). 

The analysis can be easily generalized to the case of an
arbitrary bare action for the elementary fermion. Consider, for
example,
replacing (\ref{Schi}) with
\be
\label{Schiprime}
S'_{\chi}=b\int
d^dx\;\bar\chi\big(-i\gamma^0\d_0+\gamma^0\Omega(-\D)\big)\chi\;,
\ee
where $\Omega(-\D)$ is a function of the spatial Laplacian 
$\D\equiv\d_i\d_i$. 
This action violates the charge conjugation and $CPT$.
In the
absence of coupling to the CFT it describes a non
relativistic fermion with dispersion relation
\be
\label{dispbare}
\omega=\Omega({\bf k}^2)\;,
\ee 
whose spin degree of freedom is decoupled from the translational
motion. In $d=4$ this is a fermion
with two states corresponding to the spin projections $\pm 1/2$ on an
arbitrary 
axis.
However, the situation changes qualitatively once we couple it to the
bulk theory. The expression (\ref{Schiprime}) translates into a
boundary term on the UV brane,
\be
\label{SUVprime}
S'_{UV}=b\int_{u=l} 
d^dx\;\bar\psi_-\big(-i\gamma^0\d_0+\gamma^0\Omega(-\D)\big)\psi_-\;,
\ee
which leads to the effective action for the light mode,
\be
\label{Seffprime}
S_{eff}'=-i\int d^dx\;\bigg[\bar{\hat\chi}_{1}\gamma^\m\d_\m\hat\chi_1
+(1-2Ml)\frac{b}{l}\bigg(\frac{l}{L}\bigg)^{1-2Ml}
\bar{\hat\chi}_1\gamma^0\big(\d_0+i\Omega(-\D)\big)\hat\chi_1\bigg]\;.
\ee
The non-relativistic term in the action is again suppressed 
relative to the LI kinetic term induced from the bulk. The
corresponding dispersion relation reads,
\be
\label{disprel22}
\omega=\pm
|{\bf k}|\bigg[1-(1-2Ml)\frac{b}{l}\bigg(\frac{l}{L}\bigg)^{1-2Ml}\bigg]
+\Omega({\bf k}^2)(1-2Ml)\frac{b}{l}\bigg(\frac{l}{L}\bigg)^{1-2Ml}\;,
\ee
and the positive-frequency eigenspinor $\chi_1$ satisfies the relation
(\ref{BCUV3}) implying that its polarization is aligned with the
momentum. Note that in this setup $CPT$ emerges\footnote{If $\Omega(0)$ is non-zero, there is a small violation of
  $CPT$ for modes with very low 
  momentum, which can be interpreted as the
  presence of a chemical
  potential of order $\Omega(0)(l/L)^{1-2Ml}$. } together with
LI.

\section{Emergent chirality from Lifshitz flows}
\label{sec:4}

\subsection{The domain wall geometry}
\label{ssec:dw}

Lifshitz space introduced in \cite{LSatUV1,LSatUV2} is
a useful tool for holographic studies of non--relativistic
theories. It possesses the isometry 	
\begin{align} 
\label{LifshitzIsometry}
&t \mapsto \lambda^z t,~~~~ x_i \mapsto \lambda x_i,~~i=1,\ldots,d-1\\
&u \mapsto \lambda u,
\end{align} 
where $\lambda$ is a scaling parameter. The AdS space-time is a
special case of the Lifshitz geometry with $z=1$. By extrapolating the
ideas of holography to other values of $z$ one expects the physics of
Lifshitz space-time to capture the properties of  
strongly coupled $d$-dimensional non-relativistic theories
invariant under the anisotropic scaling of time and space
(\ref{LifshitzIsometry}). Embeddings of Lifshitz solutions into
supergravity and string theory have been discussed in 
\cite{Balasubramanian:2010uk,Donos:2010tu,Gregory:2010gx,Cassani:2011sv}.

\begin{figure}[t] 
\center{\includegraphics[width=0.45\linewidth]{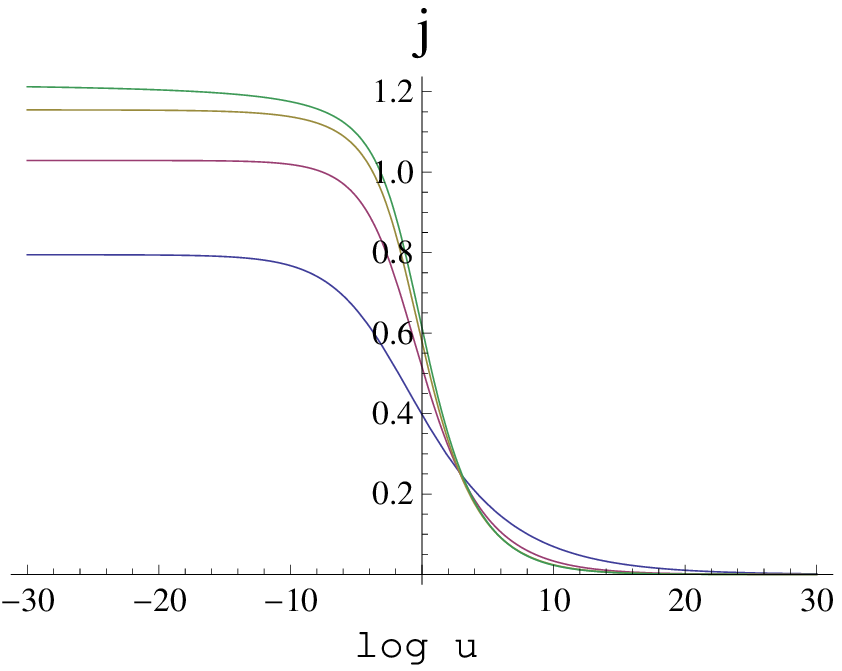}
\qquad
\includegraphics[width=0.45\linewidth]{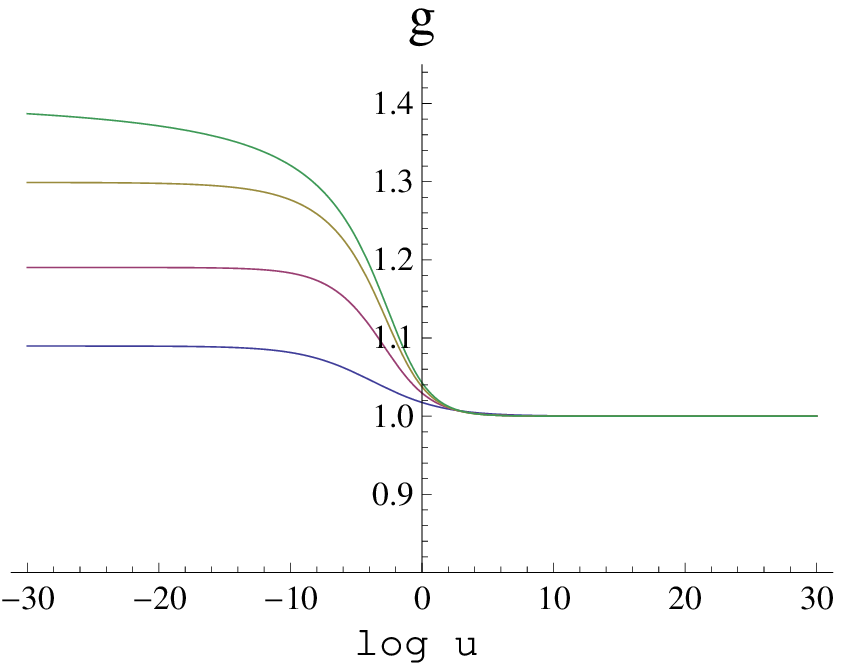}}
\caption{Functions $ j(u) $ and $ g(u) $ describing the domain 
wall solution interpolating between Lifshitz space at $u\to 0$ and AdS
at $u\to \infty$ for several values of the Lifshitz exponent and
$d=4$.
From top
to bottom:
$z=3,~2.5,~2,~1.5$.
The position $l_*$ of the domain wall has been set
to unity.
\label{FuncBeh}}
\end{figure}

To describe an RG flow from a Lifshitz UV fixed point to an IR theory
with emergent LI, we need a geometry interpolating between Lifshitz
space at small values of the holographic coordinate, $u\to 0$, and AdS
at   
$u \to \infty $. 
Geometries with these properties 
can be obtained as solutions to
the Einstein equations with matter represented by a massive vector
field $V_M$. The
action reads 	
\begin{equation}
S_{V}=\frac{1}{16\pi \kappa}\int d^{d+1}x \left(  R-2\Lambda_c
  -\frac{1}{4}F_{MN}F^{MN}-\frac{M_V^2}{2}V_NV^N \right), 
\end{equation}
where $ \Lambda_c<0 $ is a cosmological constant and
$F_{MN}=\partial_M V_N - \partial_N V_M$ is the field strength. 
A detailed study of this system was performed in \cite{LSatUV3} (see
also \cite{Bednik:2013nxa}). It was shown that it has a family of
solutions labeled by a parameter $l_*$, that are
static and invariant under 
$SO(d-1)$ rotations of the spatial coordinates. The metric and the
vector field have the form,
\begin{equation} 
\label{funcBeh}
ds^2 = \left(  \frac{l}{u} \right)^2 (-f^2(u)dt^2+dx_idx_i+g^2(u)du^2), ~~~ 
V_{t}=\frac{2}{M_Vu}f(u)j(u), ~~~V_i=V_u=0\;,
\end{equation}
where 
\be
l^2=-\frac{d(d-1)}{2\Lambda_c}\;.
\ee
The functions $f$, $g$, $j$ depend on $u$ through the combination $u/l_*$
and can be found numerically. 
They have the asymptotics
\bseq
\label{asympt}
\begin{align}
\label{asympt1}
&f=f_0(u/l_*)^{1-z}~,~~ g=g_0~,~~j=j_0~, &u \rightarrow 0\;, \\
\label{asympt2}
&f=1+f_{\infty}(u/l_*)^{-2\alpha_V}~,~~  
g=1+g_{\infty}(u/l_*)^{-2\alpha_V}~,~~
j=j_{\infty}(u/l_*)^{-\alpha_V}\;, &u \rightarrow \infty \;.
\end{align}
\eseq
Here $f_{0,\infty}$,  $g_{0,\infty}$,  $j_{0,\infty}$ are constants of
order one and
\be
\label{alphaA} 
\alpha_V = -\frac{d}{2}+\sqrt{\left( \frac{d}{2}-1 \right) ^2 + \left(
    M_Vl \right) ^2}\;.
\ee
For a given $l_*$ the solution has the form of a domain wall
interpolating between the Lifshitz and AdS spaces and centered at
$u=l_*$. 
In Fig.~\ref{FuncBeh} we plot the functions $j(u)$ and $g(u)$ for the
case $d=4$ and several values of the Lifshitz exponent. 
Numerical analysis shows
 that all three functions $ f,~g,~j$ are 
monotonically decreasing with $u$, so that the coefficients  $
f_{\infty},~g_{\infty},~j_{\infty}$ in (\ref{asympt2})
are  positive. 

The domain wall solutions exist for the vector field masses in the range
\be
\label{MAlimits}
d-1\leq (M_Vl)^2\leq\frac{d(d-1)^2}{3d-4}\;.
\ee
The 
Lifshitz exponent $z$ is related to the space-time
dimensionality $d$ and the quantity $M_Vl$. We will not need the
explicit expression, which can be found in
\cite{Bednik:2013nxa}, 
and just note
that for all values of the vector field mass satisfying
(\ref{MAlimits})
the exponent lies in the range $1\leq z\leq
d-1$.
Also, due to the upper bound (\ref{MAlimits}) on $M_Vl$ the exponent
$\alpha_V$ is bounded from above and numerically turns out to be quite
small for interesting values of $d$ (for example, for $d=3$ and $d=4$ the
upper limit on $\alpha_V$ is 0.13 and 0.35 respectively). 

In the dual picture the position of the domain wall sets the energy
scale $\Lambda_*=l_*^{-1}$ where the RG flow makes transition between
the vicinities of the Lifshitz UV theory and the LI IR fixed
point. The approach to the latter is governed by the vector operator 
${\cal O}_V^\m$ dual to the bulk
vector field. The dimension of this operator in the IR theory is
found using the standard rules of AdS/CFT,  
\be
\label{dimOA}
\dim {\cal O}_V^\m=d+\a_V\;.
\ee
Thus, the vector operator is irrelevant and the IR fixed point is
attractive as long as 
$\a_V>0$. From (\ref{alphaA}) we see that this condition coincides with
the lower bound on the vector field
mass in (\ref{MAlimits}).

\subsection{Fermions in the Lifshitz flow}

We now add fermions to the above setup. 
We consider the action, 
\begin{equation}
\label{actLif}
S = -\int d^{d+1}x \sqrt{|g|}\; i \bar{\psi} \big(\slashed D -
\xi \slashed D^{(V)}- 
M\big)\psi ,
\end{equation}
where the Dirac operator and covariant derivatives are given by
(\ref{dirac}) and we have introduced a direct coupling of the fermion
to the vector 
field via the operator\footnote{We do not consider a coupling without
  derivatives 
$\bar\psi V_N e^N_A\Gamma^A\psi$ which can be forbidden by the
symmetry $V_N\mapsto -V_N$.}
\be
\label{DiracV}
\slashed D^{(V)}\equiv\frac{(M_Vl)^2}{4} V_{N}V_{M}e^{M}_{A}\Gamma^{A}D^{N}\;.
\ee
It modifies the effective metric felt by the fermion in non-zero
vector field background,
\be
\label{effmetric}
g_{MN}\mapsto g_{MN}-\xi\frac{(M_Vl)^2}{4} V_{N}V_{M}\;.
\ee
The coupling constant $\xi$ can be different for different fermion
species and serves to probe species--dependent properties of the RG
flow.  
Note that we have omitted the boundary term $S_\d$. In contrast to 
Sec.~\ref{sec:3},  
we will concentrate here on the behavior of the dual field theory without sources
and correspondingly impose vanishing boundary conditions on
$\psi$ at $u\to 0$. In this case the boundary term vanishes together
with its variation. 

Taking a diagonal vielbein and computing the non-vanishing
components of the spin connections,
\be
\label{Lifspin}
\omega_{u00}=\frac{f'}{g}-\frac{f}{ug}~,~~~
\omega_{uij}=\frac{\delta_{ij}}{ug}
\ee 
one obtains the expressions for the operators $\slashed D$ and 
$\slashed D^{(V)}$ in the background (\ref{funcBeh}),
\bseq
\label{Lifoperators}
\begin{align}
\label{LifDirac}
&\slashed D=\frac{u}{lg}
\bigg[\d_u+\frac{1}{2}\bigg(\frac{f'}{f}-\frac{d}{u}\bigg)\bigg]\Gamma^u
+\frac{u}{lf}\Gamma^0\d_0+\frac{u}{l}\Gamma^i\d_i\;,\\
\label{LifAcouple}
&\slashed D^{(V)}=
-j^2\bigg[\frac{u}{lf}\Gamma^0\d_0+\frac{u}{2lg}
\bigg(\frac{f'}{f}-\frac{1}{u}\bigg)\Gamma^u\bigg]\;.
\end{align}
\eseq
Decomposition of the spinor into eigenstates of
$\Gamma^u$ and Fourier transform along the $(t,{\bf x})$ coordinates 
yield the system of equations,
\be
\label{Lifeqs}
\pm\big(\d_u+F_\pm(u)\big)\psi_\pm+ig\big(-\omega\gamma^0+k_i\gamma^i)\psi_{\mp}
=-i\omega G(u)\gamma^0\psi_{\mp}\;,
\ee
where
\be
\label{Liffunc}
F_{\pm}(u)=\frac{1}{2}\bigg[\frac{(1+\xi j^2)f'}{f}-\frac{d+\xi
  j^2}{u}\bigg]\mp \frac{Mlg}{u}~,~~~~~
G(u)=
g\bigg(1-\frac{1+\xi j^2}{f}\bigg)\;.
\ee
Note that the l.h.s. of (\ref{Lifeqs}) contains only LI operators,
while all Lorentz violating contributions have been grouped 
on the r.h.s.

To obtain a discrete spectrum of excitations, we cut the space-time by
an IR brane\footnote{The brane energy-momentum tensor required for a
  static solution of Einstein's equations is composed of a negative
  tension and a contribution satisfying the null energy condition
  \cite{Bednik:2013nxa}.} placed at 
$u=L\gg l_*$.
As discussed in Sec.~\ref{sec:2}, in the relativistic case there is a
massless fermion mode for the choice of boundary conditions ${\cal
  B}_-$ (\ref{Bminus}). Let us show that this choice also gives rise to a
gapless mode in the Lifshitz flow geometry. Consider first the limit
of vanishing frequency and momentum. Setting
$\omega=k_i=0$ in (\ref{Lifeqs}) we obtain the solution,
\be
\label{zerozero}
\psi_+^{(0)}=\chi_0\exp \int_u^LF_+(u')du'~,~~~~
\psi_-^{(0)}=0\;,
\ee
where $\chi_0$ is a constant chiral spinor satisfying\footnote{Recall
  that
for our choice of the bulk $\Gamma$-matrices
  $\Gamma^u=\gamma^{d+1}$, see (\ref{gammaeven}).}
$\gamma^{d+1}\chi_0=\chi_0$. This solution behaves as 
$$\psi_+^{(0)}\propto (u/L)^{Ml+d/2}$$ 
in the AdS region $u\gg l_*$. Taking into account the AdS measure 
$\int du\, u^{-d}$ entering into the mode normalization, we see that
the normalization integral for (\ref{zerozero}) is saturated at $u\sim
L$ if $Ml>-1/2$. In other words, the mode (\ref{zerozero}) is
suppressed 
in the Lifshitz part of
the space-time. 
This suggests the following strategy to solve 
Eqs.~(\ref{Lifeqs}) for non-zero
momenta. As the zeroth approximation one takes (\ref{zerozero})
together with the LI dispersion relation\footnote{We focus on the
  positive-frequency mode; for
  $\omega=-|{\bf k}|$ the analysis is similar.} 
$\omega=|{\bf k}|$. This yields the constraint on the spinor amplitude, 
\be
\label{chi0plus}
\Sigma_{\bf k}\chi_0=0\;,
\ee
where $\Sigma_{\bf k}$ has been defined in (\ref{BCUV3}). Thus $\chi_0$
corresponds to a fixed helicity. The Lorentz violating
contributions are then treated as small corrections.

We write,
\be
\label{corr1}
\psi_+=\psi_+^{(0)}+\psi_+^{(1)}~,~~~
\psi_-=\psi_-^{(1)}~,~~~\omega=|{\bf k}|+\omega^{(1)}\;.
\ee
Substituting into (\ref{Lifeqs}) and keeping only terms linear in the
perturbations we obtain,
\bseq
\label{perteqs}
\begin{align}
\label{perteq1}
&(\d_u+F_+)\psi_+^{(1)}+ig\,\Sigma_{\bf k}\psi_-^{(1)}=0\;,\\
\label{perteq2}
&-(\d_u+F_-)\psi_-^{(1)}+ig\,\Sigma_{\bf k}\psi_+^{(1)}
-ig\omega^{(1)}\gamma^0\psi_+^{(0)}=-i|{\bf k}| G\gamma^0\psi_+^{(0)}\;.
\end{align}
\eseq
One multiplies the first equation by $\Sigma_{\bf k}$ and uses the
identity $\Sigma_{\bf k}^2=0$ to eliminate the second term. The
remaining equation has the solution,
\be
\label{sigmapsip}
\Sigma_{\bf k}\psi_+^{(1)}=\tilde \chi \exp\int_u^L F_+(u')du'\;,
\ee
where the constant spinor $\tilde\chi$ satisfies the
relations 
\be
\label{tildechirel}
\gamma^{d+1}\tilde \chi=-\tilde\chi~,~~~~\Sigma_{\bf k}\tilde\chi=0\;.
\ee
Substitution of (\ref{sigmapsip}) into (\ref{perteq2})
yields the solution for 
$\psi_-^{(1)}$,
\be
\psi_-^{(1)}(u)=\tilde\psi_-^{(1)}(u)\exp\int_u^L F_-(u')du'\;,
\ee
where
\be
\label{tildepsi-}
\tilde\psi_-^{(1)}(u)=\int_0^udu'\,\big[
i\big(|{\bf k}|G(u')-\omega^{(1)}g(u')\big)\gamma^0\chi_0
+ig(u')\tilde\chi
\big]
\exp\bigg[-2Ml\int_{u'}^L\frac{g(u'')du''}{u''}\bigg]\;.
\ee
In deriving the last expression we have used $F_+-F_-=-2Mlg/u$. Note
that the outer integral in (\ref{tildepsi-}) is taken from $u=0$ to
ensure vanishing of the field on the UV boundary. 

We now impose the
boundary condition ${\cal B}_-$ on the IR brane,
\be
\label{bcLif}
\tilde\psi_-(L)=0\;.
\ee
Multiplying (\ref{tildepsi-}) successively by $\Sigma_{\bf k}$ and 
$\tilde\Sigma_{\bf k}\equiv |{\bf k}|\gamma^0+k_i\gamma^i$ and using the
relations, 
\be
\Sigma_{\bf k}\gamma^0\chi_0=2|{\bf k}|\chi_0~,~~~
\tilde\Sigma_{\bf k}\gamma^0\chi_0=0~,~~~
\tilde\Sigma_{\bf k}\tilde\chi=2|{\bf k}|\gamma^0\tilde\chi\;.
\ee
we find that the spinor $\tilde\chi$ must vanish,
\be
\tilde\chi=0\;,
\ee
whereas  
the correction to the dispersion relation reads,
\be
\label{corrdisp}
\omega^{(1)}=|{\bf k}|\;\frac{\int_0^Ldu\,G(u)
\exp\Big[-2Ml\int_u^L\frac{g(u')du'}{u'}\Big]}
{\int_0^Ldu\,g(u)
\exp\Big[-2Ml\int_u^L\frac{g(u')du'}{u'}\Big]}\;.
\ee
Finally, the correction $\psi_{+}^{(1)}$ is found from (\ref{perteq1}),
\be
\label{psi+corr}
\psi^{(1)}_+=\tilde\psi_+^{(1)}(u)\exp\int_u^LF_+(u')du'
\ee
with
\be
\label{tildepsi+}
\tilde\psi_+^{(1)}(u)=-i\int_u^L du'
g(u')\Sigma_{\bf k}
\tilde\psi_-^{(1)}(u')\exp\bigg[2Ml\int_{u'}^L\frac{g(u'')du''}{u''}\bigg]\;.
\ee
Note that the corrections preserve the property that  
the fermion wavefunction on
the IR brane $\psi_+(L)$ satisfies $\Sigma_{\bf k}\psi_+(L)=0$ implying that
it
describes a particle of fixed helicity.

Our next task is 
to infer the scaling of $\omega^{(1)}$ with the IR cutoff $L^{-1}$.
We notice that for $Ml>-1/2$
the integral in the denominator of (\ref{corrdisp}) is saturated in
the AdS region where $g(u)$ can be replaced by one. This yields,
\be
\label{den}
\int_0^Ldu\, g(u) \exp\bigg[-2Ml\int_u^L\frac{g(u')du'}{u'}\bigg]
=\frac{L}{1+2Ml}\;.
\ee
To evaluate the numerator we recall 
that the functions $f,g,j$ depend on $u$ only through
the combination $z\equiv u/l_*$ (see Sec.~\ref{ssec:dw}). 
Switching to the latter variable inside the integrals we have,
\be
\label{calcul}
\begin{split}
\int_0^Ldu\, G(u) \exp\bigg[-2Ml\int_u^L&\frac{g(u')du'}{u'}\bigg]
=l_*\int_0^{L/l_*}
dz\,G(z)\exp\bigg[-2Ml\int_z^{L/l_*}\frac{g(z')dz'}{z'}\bigg]\\ 
&=l_*\!\bigg\{
\!\int_0^{z_0}\!\!\!dz\,G(z)\exp\bigg[\!\!-2Ml\!\int_z^{z_0}
\!\frac{g(z')dz'}{z'}
-2Ml\!\int_{z_0}^{L/l_*}\!\frac{g(z')dz'}{z'}\bigg]
\\
&\qquad
+\int_{z_0}^{L/l_*}\!\!\!
dz\,G(z)\exp\bigg[\!\!-2Ml\!\int_{z}^{L/l_*}
\!\frac{g(z')dz'}{z'}\bigg]\bigg\}\\ 
&=l_*\bigg\{a_1\bigg(\frac{l_*}{L}\bigg)^{2\a_V-1}
+a_2\bigg(\frac{l_*}{L}\bigg)^{2Ml}\bigg\}\;,
\end{split}
\ee
where the coefficients
\bseq
\label{cs}
\begin{align}
\label{c1}
&a_1=\frac{f_\infty-\xi j^2_\infty}{2Ml-2\a_V+1}\;,
\\
\label{c2}
&a_2=
z_0^{2Ml}\int_0^{z_0}\!\!\!dz\,G(z)
\exp\bigg[-2Ml\int_z^{z_0}\frac{g(z')dz'}{z'}\bigg]
-\frac{(f_\infty-\xi j^2_\infty)z_0^{2Ml-2\a_V+1}}{2Ml-2\a_V+1}\;
\end{align}
\eseq
are independent of $l_*$ and $L$. On the other hand, they depend on
the direct coupling $\xi$ between the fermion and the vector field and
thus can be different for different fermion species.
In the second line of (\ref{calcul}) we divided the integral in two
parts with the intermediate point $z_0$ lying in the range $1\ll
z_0\ll L/l_*$ and then used the asymptotic form of the metric
functions (\ref{asympt2}) at $z>z_0$. Of course, 
the expression in the last line of (\ref{calcul}) 
is independent of the choice of $z_0$. 
Combining (\ref{calcul}) and (\ref{den}) we obtain
\be
\label{corrdisp1}
\omega^{(1)}=|{\bf k}|\,\bigg[a_1(1+2Ml)\bigg(\frac{l_*}{L}\bigg)^{2\a_V}+
a_2(1+2Ml)\bigg(\frac{l_*}{L}\bigg)^{1+2Ml}
\bigg]\;.
\ee
This gives power-suppressed correction to the fermion velocity as
long as $\a_V>0$, $Ml>-1/2$. To sum up,
similar to the model of Sec.~\ref{sec:3}, we have found that the
gapless fermion mode has an almost LI dispersion relation with the
number and structure of independent degrees of freedom matching those
of a Weyl fermion in $d$ dimensions (a single helicity component for
$d=4$). Clearly, this mode is described by an approximately LI effective action
with the fermion field transforming in the chiral representation of the
Lorentz group.

The form of corrections (\ref{corrdisp1}) to the relativistic dispersion
relation has a natural interpretation in the dual field theory. At the
IR fixed point the theory is described by a CFT containing 
a spinor operator ${\cal O}_\psi$ and a 
vector operator ${\cal O}_V^\m$ with dimensions
(\ref{dimO}) and (\ref{dimOA}).
The RG flow corresponds to deforming the CFT action 
by an irrelevant Lorentz violating perturbation,
\begin{equation}
\label{Ldef}
\delta S_{CFT}=\int d^dx \big(c_1\; l_*^{\a_V}
\mathcal{O}_V^0 + c_2\;
l_*^{1+2Ml}\bar{\mathcal{O}}_\psi \gamma^0\d_0
  \mathcal{O}_\psi\big)\;, 
\end{equation}
where $c_{1,2}$ are some dimensionless constants.
These two terms generate the two contributions in
(\ref{corrdisp1}) after the theory enters into the confining phase. 
Note that a double insertion of the deformation ${\cal O}_V^0$ is
required to affect the
fermion dispersion relation, as in the unperturbed CFT the
three-point function ${\cal O}_\psi\bar{\cal O}_\psi {\cal O}_V$
vanishes due to the symmetry of the bulk action under $V_M\mapsto
-V_M$.  

Finally, let us comment on the domain of validity of the 
calculation leading to (\ref{corrdisp1}). A necessary requirement is
that the correction (\ref{psi+corr}) 
to the fermion wavefunction is smaller than the
wavefunction at the zeroth order (\ref{zerozero}). 
From the estimates
\be
\psi_+^{(1)}/\psi_+^{(0)}\sim kL\tilde\psi_-^{(0)}/\chi_0
\sim kL^2\omega^{(1)}
\ee
one obtains the condition
\begin{equation} 
\label{kupper}
|{\bf k}| \ll L^{-1}\min\big\{(L/l_*)^{\a_V},(L/l_*)^{1/2+Ml}\big\}\;.
\end{equation}
Alternatively, the same condition can be derived from the
requirement that the correction to the gapless mode energy $\omega^{(1)}$
is smaller than the splitting between the energies of the massless
mode
and the next eigenstate
of the unperturbed relativistic Dirac operator. The latter eigenstate 
has relativistic
dispersion relation with the mass of
order $1/L$, so the splitting is of order $(|{\bf k}|L^2)^{-1}$.
The upper limit (\ref{kupper}) on the momentum is parametrically larger
than the gap $1/L$ between the gapless mode and the next bound state. It
may even seem that the perturbative expansion 
can be pushed to momenta higher than the
Lorentz violation scale $1/l_*$ if $\a_A>1$ and $Ml>1/2$. However, we expect 
that the structure of higher 
order terms in the expansion will restrict its domain of validity to
$|{\bf k}|\ll 1/l_*$, similar to the case of scalar theory
\cite{Bednik:2013nxa}. Anyway, this issue is irrelevant for the
setup studied in this section: as pointed in Sec.~\ref{ssec:dw},
for the interesting values of space-time dimensionality
the exponent $\a_V$ is
smaller than 1. 

\section{RG flows with fermions in odd dimensions}
\label{app:1}

In this section we consider the holographic RG flow in the case when
the space-time dimensionality $d$ of the dual field theory is
odd. This affects the form of the bulk $\Gamma$-matrices. Instead of
(\ref{gammaeven}) we choose,
\be
\label{gammaodd}
\Gamma^\m=\begin{pmatrix}
0&\gamma^\m\\
\gamma^\m& 0
\end{pmatrix}~,~~~\m=0,\ldots,d-1,
\qquad
\Gamma^u=\begin{pmatrix}
-1&0\\
0& 1
\end{pmatrix}\;,
\ee
where $\gamma^\m$ are the $d$-dimensional Dirac matrices. Then the
analysis of
Secs.~\ref{sec:3}, \ref{sec:4} goes through with minor changes. One
concludes that the two-brane model with the action (\ref{SUV}) or
(\ref{SUVprime}) on the UV brane and ${\cal B}_+$ boundary conditions
on the IR brane gives rise to a gapless fermion mode with almost LI
dispersion relation. The same holds for the case of Lifshitz domain
wall with ${\cal B}_-$ boundary conditions in the IR. Of course, there
is no notion of chirality in odd dimensions and the fermions transform
in the Dirac representation of $SO(d-1,1)$. The absence of the gap is
protected by spatial parity which forbids a LI mass term for fermions
in odd dimensions (see e.g. \cite{Appelquist:1986qw}). Indeed,
under the reflection of a single
axis a fermion in odd dimensions transforms as,
$$
\chi(x^0,x^1,x^2,\dots)\mapsto \gamma^1\chi(x^0,-x^1,x^2,\ldots)\;.
$$
It is straightforward to check that a mass term $m\bar\chi\chi$ changes sign
under this transformation. On the other hand, the original bulk
action (\ref{BulkAction}) or (\ref{actLif}) respects parity which
in $(d+1)$ even dimensions can be defined as
$$
\psi(x^0,x^1,x^2,\dots)\mapsto \Gamma^{d+2}\Gamma^1\psi(x^0,-x^1,x^2,\ldots)\;,
$$
where $\Gamma^{d+2}=i\Gamma^0\Gamma^1\ldots\Gamma^{d-1}\Gamma^u$.
 
A new feature that appears in the case of odd $d$ is a possible
mismatch between the number of degrees of freedom of non-relativistic
and relativistic fermions. Namely, the spinor representation of the group
$SO(d-1)$ of spatial rotations can be decomposed into the left and
right parts, each having dimension $2^{(d-3)/2}$, which is thus the
minimal number of components of a non-relativistic spinor. This does
not fit into the Dirac representation of $SO(d-1,1)$ which has
$2^{(d-1)/2}$ components. Therefore, one does not expect emergence of LI
in systems with minimal non-relativistic fermions. Let us analyze this
question explicitly. We take $d=3$ and consider the holographic setup
with two branes similar to that studied in Sec.~\ref{sec:3}. First, we
describe the model from the 3-dimensional viewpoint. The 
spinor operator ${\cal O}_\psi$ of the dual CFT 
can be decomposed
into the upper and lower components,
\be
\label{odecomp}
{\cal O}_\psi=\begin{pmatrix}
{\cal O}_U\\
{\cal O}_D
\end{pmatrix}\;.
\ee
Taking the $\gamma$-matrices in the form,
\be
\label{3dgammas}
\gamma^0=i\sigma_3~,~~~\gamma^1=\sigma_1~,~~~\gamma^2=\sigma_2\;,
\ee
where $\sigma_i$ are the Pauli matrices, we observe that ${\cal O}_U$
and ${\cal O}_D$ do not mix under purely spatial rotations. Hence, it
is consistent with spatial isotropy to couple only one of them, say
${\cal O}_U$, to an external dynamical field. This leads to the action,  
\be
\label{3dact}
S=S_{CFT}+l^{Ml}\int d^3x\, \chi_U^\dagger{\cal O}_U
+b\int d^3x\,
\chi_U^\dagger\big(i\d_0-\Omega(-\D)\big)\chi_U\;,
\ee  
where $\chi_U$ is a single-component minimal $SO(2)$ spinor and
$\Omega(-\D)$ is an arbitrary function of the spatial Laplacian. Note that
this action breaks the discrete symmetries $C$, $P$, $T$ and $CPT$ as
all of them
interchange ${\cal O}_U$ and ${\cal O}_D$.

The holographic description of this
model is obtained by making only the upper component of the bulk
spinor $\psi_-$ on the UV brane dynamical with the action,
\be
\label{SUV3d}
S_{UV}=b\int_{u=l} d^3x 
\;\psi^\dagger_{-,U}\big(i\d_0-\Omega(-\D)\big)\psi_{-,U}\;,
\ee
whereas imposing the Dirichlet boundary condition on the lower
component, 
\be
\label{zeroD}
\psi_{-,D}\big|_{u=l}=0\;.
\ee
From (\ref{SUV3d}), (\ref{BulkAction}) we obtain the second boundary
condition on the UV brane,
\be
\label{3dU}
\big[b\big(i\d_0-\Omega(-\D)\big)\psi_{-,U}+\psi_{+,U}\big]\big|_{u=l}=0\;.
\ee
Combining with ${\cal B}_+$ boundary conditions on the IR brane and
substituting the bulk solution (\ref{psibulk}), (\ref{functions}) into
(\ref{zeroD}), (\ref{3dU}) yields,
\be
\label{3dU1}
\big[mb\big(\omega-\Omega(k^2)\big)f_-(l)+
\omega f_+(l)\big]\chi_{1,U}=0\;.
\ee
Expansion of the functions $f_{\pm}(l)$
at small argument gives the dispersion relation of the
gapless mode,
\be
\label{3ddisp}
\omega=\Omega(k^2) (1-2Ml)\frac{b}{l}\bigg(\frac{l}{L}\bigg)^{1-2Ml}\;.
\ee
Clearly, this is not relativistic. In the infrared limit $l/L\to 0$
the dispersion relation becomes degenerate, $\omega=0$, implying that,
the excitation of this mode does not cost any energy. In other words, the 
system contains infinite number of states with zero energy 
different from the vacuum. This situation appears rather
pathological. It is excluded if we impose from the beginning one of
the symmetries $C$,
$P$, $T$ or $CPT$.

\section{Conclusions}
\label{sec:5}

In this paper we have studied emergence of LI in strongly coupled
field theories with fermions. We considered two models. In the first
model a strongly interacting relativistic CFT is coupled at a high
energy scale to an elementary fermion with non-relativistic
action. This generates an RG flow exhibiting LI in the IR. The second
setup describes a field theory flowing from a UV fixed point with
Lifshitz scaling to a LI IR fixed point. In both cases we assumed that
the theories enter into a confinement phase at low energies, which
gives rise to a discrete spectrum of excitations. We used the holographic
duality to describe the strong coupling. 

In the main part of the paper we focused on field theories
living in even-dimensional space-time and analyzed how the notion of
chirality arises together with emergent LI. We have found that if 
the theory possesses low-lying fermionic excitations, the
latter are described by approximately LI effective actions with the
fermion wavefunction transforming in the Weyl representation of the
Lorentz group. The Lorentz violating corrections to the effective
action are power-law suppressed by the ratio between the IR
confinement scale and the UV scale of Lorentz violation. The exponent
in the power-law is related to the dimension of the Least Irrelevant
Lorentz Violating Operator (LILVO) in the IR theory, 
in complete analogy with the
scalar case studied in \cite{Bednik:2013nxa}. A kinematic property
that appears essential for such behavior is the coincidence between
the number of independent components of the non-relativistic fermion
operator in the UV theory and the relativistic Weyl spinor.

In the last section we have addressed the case of odd space-time
dimensionality. We observed that the setups preserving the kinematic
match mentioned above still lead to low-energy theories with
approximately relativistic fermions which are protected from acquiring
a mass by spatial parity. We also considered a model that violates the
above match and showed that in this case the low-lying fermion mode
does not exhibit LI. We pointed out that this situation is excluded if
the theory respects one of the discrete symmetries $C$, $P$, $T$ or
$CPT$. This is one more manifestation of the deep role played by
these symmetries in the phenomenon of emergent LI.

From the phenomenological perspective, our work represents a step
towards implementing the idea of emergent LI in a realistic particle
physics setting. Of course, an important open issue on this way is
inclusion of gauge fields. As discussed in \cite{Bednik:2013nxa}, 
the holographic
description of strongly coupled RG flows with gauge fields will 
presumably require considering a more complicated bulk sector
containing an interaction of the gauge fields with a dilaton. More
generally, one can
ask whether the gauge symmetry can be an emergent property
appearing together with emergent LI.

Another interesting question concerns the minimal value
of LV couplings predicted in this framework. We
have demonstrated that these couplings are suppressed by a power-law
factor  
$(\Lambda_{IR}/ \Lambda_* )^\a$, where the ratio $\Lambda_{IR}/
\Lambda_*$ characterizes the ``duration'' of the RG flow leading to
the emergence of LI. Clearly, this duration cannot be arbitrarily
large. In the models inspired by quantum gravity the natural value for
$\Lambda_*$ is at or below the Planck scale. In particular, in the
context of the Horava's proposal, $\Lambda_*$ should be less than
$10^{15}$ GeV \cite{Blas:2009ck,Blas:2010hb}. On the other hand, the
scale $\Lambda_{IR}$ is bounded below by particle physics
experiments. As mentioned in the Introduction, $\Lambda_{IR}$ sets the
scale of compositeness for the SM fields. Direct searches for excited
fermionic states\footnote{More stringent limits
can come from the physics of flavor, but they are model
dependent.} put a lower bound on $\Lambda_{IR}$ at the level of a
few TeV \cite{Aad:2013jja,Khachatryan:2014aka}.  Taking for the estimate $\Lambda_{IR}\gtrsim
10$~TeV, we find that the duration of the RG flow is bounded from
below by $\Lambda_{IR}/\Lambda_*\gtrsim 10^{-11}$. In all 
holographic examples considered in
Ref.~\cite{Bednik:2013nxa} and this paper the exponent $\a$ in the
suppression has been found to be smaller than 2. Thus, the minimal
size of LV operators predicted by these models is of order $10^{-22}$,
which is marginally compatible with the existing bounds
\cite{Kostelecky:2008ts}. In this light it  
will be important to 
understand whether the condition $\a<2$ is universal or there exist RG
flows with emergent LI that avoid~it.

To conclude, the scenario of emergent LI provides an interesting 
interplay between
high-energy particle physics and precision tests of relativity. An
improvement of the experimental lower limits on the compositeness
scale of SM fields and / or tightening of the constraints on LV
parameters will be able to test this idea or rule it out.

\paragraph{Acknowledgments} We thank Roberto Contino, Mohamed Anber
and Oriol Pujolas for fruitful 
discussions. We are grateful to Anatoly
Dymarsky for correspondence and to Diego Blas for useful comments on the draft. 
The work of I.K. is supported by the RFBR grant 14-02-31429,
S.S. is supported by the Swiss National Science Foundation.

\end{document}